\documentclass[sigconf, nonacm, 9pt, natbib=false]{acmart} %
\AtBeginDocument{%
  }

\copyrightyear{}
\acmYear{}
\acmDOI{}

\acmISBN{}

\usepackage{hyperref}

\newtheorem{example}{Example}
\usepackage{graphicx}
\usepackage{subcaption}   %
\usepackage{adjustbox} 
\usepackage{multirow}
\usepackage{csquotes}

\usepackage[style=numeric-comp,sorting=none]{biblatex}
\AtEveryBibitem{
    \clearname{editor}
    \clearfield{series}
    \clearfield{isbn}
    \clearfield{issn}
    \clearfield{day}
    \clearfield{month}
    \clearfield{place}
    \clearlist{location}
    \clearfield{eventtitle}
    \ifentrytype{inproceedings}{\clearlist{publisher}}{}
    \ifentrytype{software}{}{%
        \clearfield{url}%
        \clearfield{urlyear}%
    }
}

\usepackage[font=small,labelfont=bf]{caption}

\begin{document}

\title{Exploiting Movable Logical Qubits \mbox{for Lattice Surgery Compilation}}

    \author{Laura S. Herzog}
\email{laura.herzog@tum.de}
\affiliation{%
\department{Chair for Design Automation}
  \institution{Technical University Munich}
  \country{Germany}
}

\author{Lucas Berent}
\affiliation{%
\department{Chair for Design Automation}
  \institution{Technical University Munich}
  \country{Germany}
}
\affiliation{%
  \institution{Iceberg Quantum}
  \city{Sydney}
  \country{Australia}
}

\author{Aleksander Kubica}
\affiliation{%
    \institution{Yale Quantum Institute \& Department of Applied Physics}
  \institution{Yale University}
  \city{New Haven, CT}
  \country{USA}
}

\author{Robert Wille}
\affiliation{%
  \department{Chair for Design Automation}
  \institution{Technical University of Munich}
  \country{}
}
\affiliation{%
  \institution{Munich Quantum Software Company}
  \country{Germany}
}

\renewcommand{\shortauthors}{Herzog et al.}

\begin{abstract}

Lattice surgery with two-dimensional quantum error correcting codes is among the leading schemes for fault-tolerant quantum computation, motivated by superconducting hardware architectures.
In conventional lattice surgery compilation schemes, logical circuits are compiled following a place-and-route paradigm, where logical qubits remain statically fixed in space throughout the computation.
In this work, we introduce a paradigm shift by exploiting movable logical qubits via teleportation during the logical lattice surgery CNOT gate.
Focusing on lattice surgery with the color code, we propose a proof-of-concept compilation scheme that leverages this capability. %
Numerical simulations show that the proposed approach can substantially reduce the routed circuit depth compared to standard place-and-route compilation techniques.
Our results demonstrate that optimizations based on movable logical qubits are not limited to architectures with physically movable qubits, such as neutral atoms or trapped ions---they are also readily applicable to superconducting quantum hardware.
An open-source implementation of our method is available on GitHub \url{https://github.com/munich-quantum-toolkit/qecc}.

\end{abstract}

\maketitle

\begin{figure*}[th]
    \centering
    \includegraphics[width=0.87\linewidth]{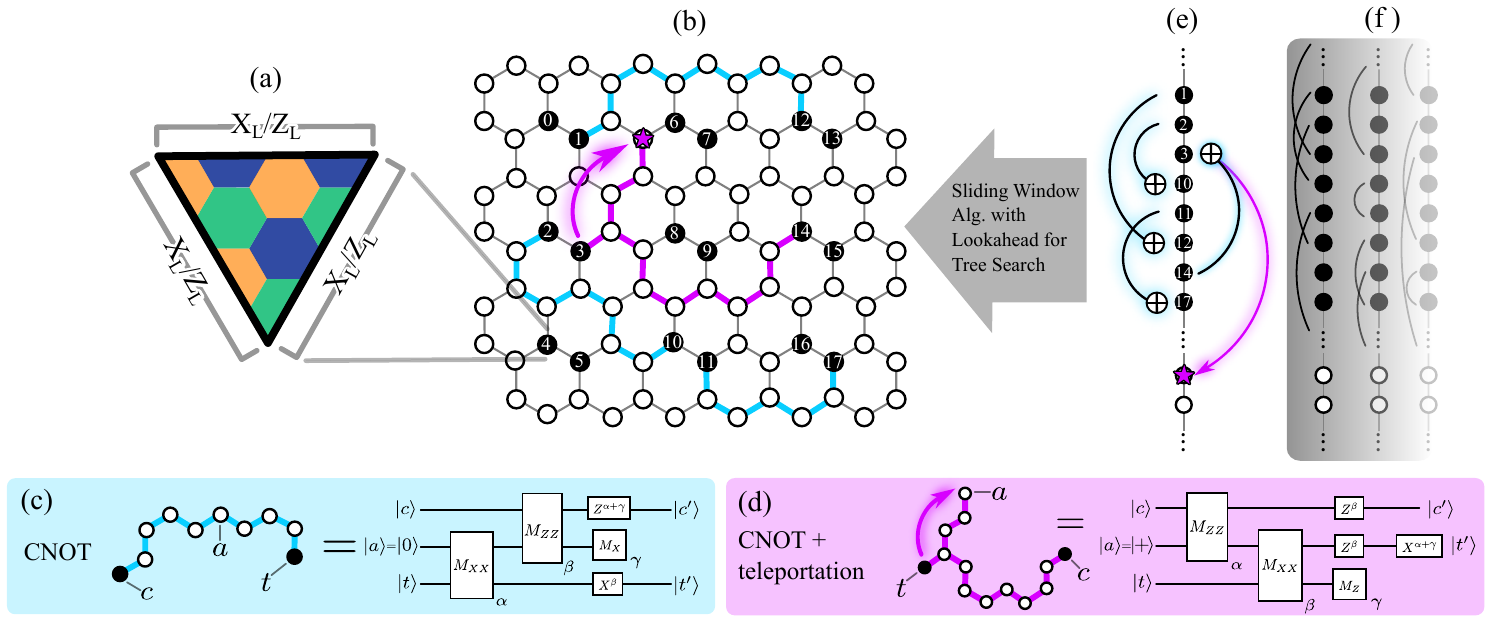}
    \caption{Overview of the movable logical qubits approach. (a) A color code patch encodes a single logical qubit which can be used as either a logical data or ancilla patch.
    (b) Macroscopic routing graph $\mathcal{R}$; black and white nodes represent data and ancilla patches, respectively.
    Paths in blue indicate standard CNOT gates performed via lattice surgery with the measurement-based scheme shown in (c); the pink tree depicts a CNOT with teleportation as shown in (d). The example displays that the target qubit with label 3 is teleported to the position marked by a pink star. (e, f) display the data and ancilla patches 
    as 1D linearized view of the routed layers. 
    (f) shows the future routed layers that determine the teleportation-based movements in (e).}  %
    \label{fig:overview}
\end{figure*}

\section{Introduction}

    \emph{Quantum error correction} (QEC) is essential for fault-tolerant quantum computation~\cite{shor1996fault, steane1999efficient, Shor1995, preskill1998reliable, kitaev1997quantum, eisert_mind_2025} and implementation of large-scale quantum algorithms~\cite{dalzell2023quantum}. 
    A major design task of scalable fault-tolerant quantum computation is efficient logical compilation. 
    In general, to compile a logical quantum circuit in a fault-tolerant manner, one has to pick a specific quantum hardware; 
    then, one chooses a suitable QEC code and maps the quantum circuit into basic logical operations on the given architecture that can be further decomposed into physical operations.
    
     This work considers planar quantum hardware with static physical qubits and local connectivity, motivated by superconducting architectures. 
     A native choice for those architectures are topological codes in two dimensions with local qubit connectivity and logical operations implemented via \textit{lattice surgery}. 
     Based on this, the compilation is performed, i.e., the translation of logical circuits to operations based on lattice surgery.
     The key aim of this task is to achieve the lowest possible space-time overheads---the product of the total number of required physical qubits and the schedule depth---of a computation using some QEC code.
    
    Most existing lattice surgery compilation approaches~\cite{molavi_dependency-aware_2024, beverland_surface_2022, herr_lattice_2017, zhu_ecmas_2023, lao_mapping_2018, watkins_high_2024, silva_multi-qubit_2024, robertson_resource_2025} assume a \textit{place-and-route} paradigm, where logical data qubits are assigned a static position at the start of the procedure (\enquote{mapping}) 
    and %
    logical entangling operations require finding paths (\enquote{routing}) between corresponding logical qubits.
    However, this assumption of static logical data qubits neglects additional degrees of freedom available through lattice surgery operations.
    In particular, measurement-based CNOT implementations, that are typically realized using lattice surgery, naturally allow movable logical data qubits via ``free'' logical qubit \emph{teleportations}---which require the same number of lattice surgery operations as the standard CNOT scheme.
    Notably, we can realize movable logical qubits in quantum architectures with static physical qubits, such as superconducting quantum hardware.
    To the best of our knowledge, this degree of freedom has not been exploited in any previous work on lattice surgery compilation.
    
    This work proposes a heuristic compilation routine that exploits movable logical data qubits via teleportations during CNOT operations and examines how this flexibility allows us to reduce the depth of the compiled schedules on a given architecture. 
    Locations of logical data qubits are dynamically adjusted during the compilation process with the goal of executing more gates in parallel.
    The proposed approach is based on two-dimensional topological codes; for concreteness we focus on a color code architecture, because the structure of the logical operators of the color code enables us to utilize movable logical qubits via teleportations in a particularly flexible and elegant way. We build upon Ref.~\cite{herzog_lattice_2025} which is the only available work in such a setup.
    The suggested measurement-based CNOT adaption is straightforwardly applicable to other topological codes, such as the surface code.
    Simulations with the proposed method identify regimes where teleportation-based movements significantly reduce routed layers, marking an initial step toward lattice surgery compilation beyond the place-and-route paradigm.
 
    This paper is structured as follows. 
    In ~\autoref{sec-background}, color codes, lattice surgery, and previous work are reviewed.
    This is followed by the main idea of exploiting movable logical qubits via the simultaneous CNOT + teleportation protocol and how this can be used to reduce routed depths in ~\autoref{sec-main-idea}. 
    Based on this, the proposed heuristic compilation routine, which makes use of the movable logical qubits via teleportations, is presented in ~\autoref{sec-heuristic}. 
    The performance of the method is studied numerically in ~\autoref{sec-numerics}. We conclude in ~\autoref{sec-conclusion}.

\section{Background and Related Work}\label{sec-background}
    QEC codes use $n$ physical qubits to encode $k<n$ logical qubits.
    Topological codes are among the most important class of QEC codes---they have favourable properties, such as geometric locality, which make them compatible with superconducting quantum hardware.
    There are various types of topological codes~\cite{bombin_introduction_2013}, including
    the \textit{surface code}~\cite{Kitaev_2003, Dennis_2002, fowler_surface_2012}, which is very broadly used due to good available decoders, its high threshold, and ease of physical implementation;
    and the \textit{color code}~\cite{bombin_topological_2006,kubica_abc_2018}, which has certain benefits compared to the surface code. For instance, transversal Clifford gates and supporting both logical $X_L$ and $Z_L$ operators along each boundary. %
    Even the decoding task---which is known to be challenging for color codes---has been shown to come closer to the performance of the surface code~\cite{Kubica2023,beverland_cost_2021,koutsioumpas_colour_2025}. 
    The color code can be implemented on a triangular patch as detailed in the following example.%
    \begin{example}
        Consider the triangular color code patch in ~\autoref{fig:overview}{a} that encodes one logical qubit. 
        Physical data qubits are placed on vertices of the lattice and both $X$ and $Z$ stabilizers are associated with the faces within the triangular patch.
        The distance is $d=5$, which determines how many errors $t=(d-1)/2$ can be corrected in principle. 
        Minimum-weight logical operator representatives $X_L$ and $Z_L$ can be found along each side of the triangular patch.
    \end{example}
    To perform lattice surgery compilation, which translates a logical circuit into operations on a 2D architecture, one abstracts away from the specific form of the color code patches and represents each of them as vertices $V_\mathcal{R}$ on a macroscopic routing graph \mbox{$\mathcal{R} = (V_\mathcal{R}, E_\mathcal{R})$} as shown in ~\autoref{fig:overview}{b}. Black nodes are logical data qubits and white nodes represent logical ancilla qubits. 
    One refers to choosing a layout of logical data and ancilla patches, and placing specific data qubit labels on the respective data patches as \textit{mapping}. 
    In order to execute logical two-qubit gates like a CNOT, the respective control and target qubits need to be connected on the macroscopic routing graph $\mathcal{R}$ following the basic connectivity given by the edges $E_\mathcal{R}$. 
    This constitutes a \textit{routing} task, where one finds many non-overlapping paths such that the final schedule is as short as possible.
    Examples of such paths for CNOT gates are displayed in blue in ~\autoref{fig:overview}{b}.  %
    Along each of these paths, a measurement-based CNOT scheme (\autoref{fig:overview}{c}) is performed. 
    Thus, a path must contain an additional logical ancilla patch, which can be placed on any vertex along the path if a color code architecture is assumed. 
    In this scheme, lattice surgery is performed for the $Z_LZ_L$ and $X_LX_L$ measurements.

    Given a layout, the main objective of lattice surgery compilation is to reduce the depth of the final schedule $\widetilde{d}$. This means, if the input logical circuit allows to perform gates in parallel, one tries to simultaneously fit as many of these parallel gates with non-overlapping paths on the macroscopic routing graph $\mathcal{R}$.
    The input logical quantum circuit is split into $d_\mathrm{L}$ \textit{logical layers}, where in each layer gates can be executed in parallel as they have disjoint logical qubit support; its depth $d_\mathrm{L}$ provides the lower bound on the routed depth.
    The logical layer structure is based on the \emph{directed acyclic graph} (DAG) of the circuit~\cite{javadiabhari2024quantumcomputingqiskit, childs-dag}. 
    Mapping and routing on some macroscopic routing graph $\mathcal{R}$ introduces further restrictions such that the resulting number $\widetilde{d}$ of \textit{routed layers} of the final schedule is likely to be higher than $d_\mathrm{L}$. 
    Hence, a good compilation scheme attempts to lower $\widetilde{d}$ and getting as close as possible to $d_\mathrm{L}$.

 In general, lattice surgery~\cite{horsman_surface_2012,fowler2019lowoverheadquantumcomputation, vuillot_code_2019} is a technique that allows to perform a CNOT gate in 2D using nearest-neighbor connectivity with the measurement-based scheme shown in \autoref{fig:overview}{c}. 
 The transversal CNOT implementation between two color code patches would require non-local connectivity otherwise. 
 Originally, lattice surgery was developed for the surface code, but it can be realized for other topological codes as well~\cite{landahl_quantum_2014}. 
 Furthermore, the technique can also be applied to more general QLDPC codes~\cite{cohen_low-overhead_2022,cowtan_ssip_2024, cross_improved_2025, baspin_fast_2025}.
    Together with single-qubit Clifford operations---which do not require code deformation and can be applied transversally in the color code---as well as T state injection, a universal gate set can be obtained.

Related work on lattice surgery compilation mostly focuses on the surface code~\cite{molavi_dependency-aware_2024, beverland_surface_2022, herr_lattice_2017, zhu_ecmas_2023, lao_mapping_2018, watkins_high_2024, silva_multi-qubit_2024, robertson_resource_2025}, leaving a largely unadressed gap in the literature for other topological codes. 
Another related but unresolved issue is whether directly compiling Clifford+T gates (as in this work) to lattice surgery operations or translating the gates into multi-qubit-Pauli measurements through Clifford conjugation first~\cite{litinski_game_2019} is to be favored, despite of early studies~\cite{leblond2025quantumresourcecomparisonleading, mcardle2025fastcuriousacceleratefaulttolerant}. 
Furthermore, one should note that with T state generation becoming cheaper~\cite{gidney_magic_2024}, %
optimizing logical Cliffords and taking into account their compilation cost become increasingly important~\cite{mcardle2025fastcuriousacceleratefaulttolerant}.

Even though not yet introduced in a compilation setup as considered in this work, there are different methods that allow to move logical information on 2D structures. Examples are standard teleportation schemes with lattice surgery~\cite{litinski_game_2019} and more time efficient schemes as for instance sliding and gliding of patches for the surface code~\cite{mcewen_relaxing_2023}. 
Note that the CNOT + teleportation utilized in this work does not impose a higher time overhead in terms of lattice surgery operations than the standard CNOT scheme. This stands in contrast to recent independent work \cite{sharma_space-time_2025} where additional time overhead is deliberately incurred to achieve denser layouts with movable logical qubits.
Finally, note that for other architectures than considered here, i.e., load/store architectures with dense memories, movement of qubits is unavoidable~\cite{kobori_lsqca_2024-1}.

\section{Main Idea}\label{sec-main-idea}
The goal of this work is to go beyond the traditional place-and-route paradigm commonly assumed in hardware architectures with static physical qubits, such as superconducting hardware.
Although physical qubits are static, logical data can be dynamically moved, enabling reductions in the routed depth during lattice surgery compilation.
In particular, we utilize movable logical qubits via teleportation steps that are implemented using the measurement-based CNOT scheme.
For this, we first show in~\autoref{sec-mainidea-micro} how the measurement-based CNOT circuit from~\autoref{fig:overview}{c} can be adapted to execute the required teleportation allowing for movable logical qubits. 
Then, ~\autoref{sec-mainidea-macro} illustrates why such CNOT + teleportation steps can be useful. 

\subsection{Movable Logical Qubits at the Microscopic Level}\label{sec-mainidea-micro}

\begin{figure}[t]
    \centering
    \begin{subfigure}[t]{0.65\linewidth}
        \centering
        \adjustbox{valign=t}{\includegraphics[width=\linewidth]{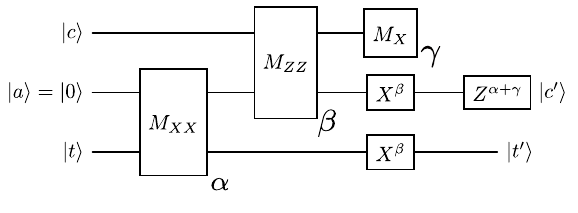}}
        \caption{}
        \label{fig:cnotmove_a}
    \end{subfigure}%
    \hfill
    \begin{subfigure}[t]{0.3\linewidth}
        \centering
        \adjustbox{valign=t}{\includegraphics[width=\linewidth]{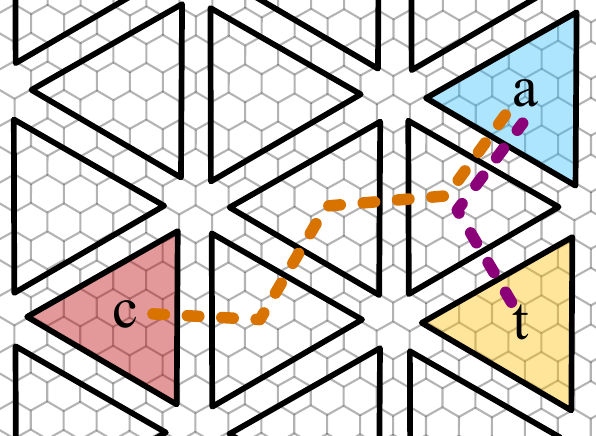}}
        \caption{}
        \label{fig:cnotmove_b}
    \end{subfigure}

    \caption{
        (a) CNOT + teleportation of control to ancilla. 
        $\alpha, \beta, \gamma \in \{0, 1\}$ denote the measurement outcomes of the corresponding Pauli measurements. (b) Possible teleportation on the geometry with the color code on the microscopic level.
        The $X_LX_L$ and $Z_LZ_L$ measurements are performed along the violet and orange dashed lines, respectively.
    }
    \label{fig:cnot+move}
\end{figure}

\begin{figure}
    \centering
    \includegraphics[width=0.65\linewidth]{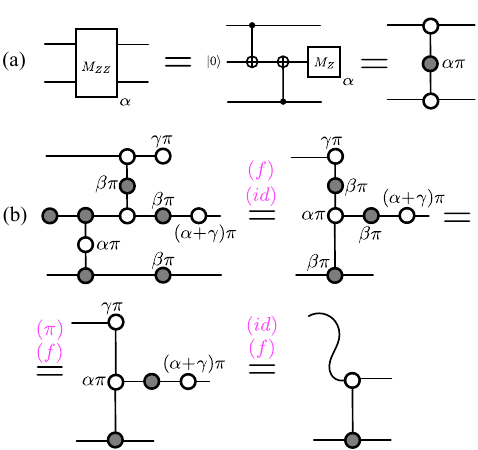}
    \caption{Derivation of CNOT + teleportation from control to ancilla with ZX calculus. Z-spiders are depicted in white and X-spiders in grey. (a) Representation of $ZZ$ measurement as quantum circuit and corresponding ZX diagram. (b) Translation of circuit in ~\autoref{fig:cnotmove_a} into ZX diagrams. The applied ZX rules per step are displayed in pink following~\cite{wetering_zx-calculus_2020}. The final expression is topologically equivalent to the CNOT gate in ZX calculus.}%
    \label{fig:zxderivation}
\end{figure}

Quantum computing with lattice surgery allows us to perform ``free'' movements of logical information via teleportation. 
To see this, one can adapt the measurement-based CNOT scheme from ~\autoref{fig:overview}{c} by applying a measurement on the target patch, such that the target patch is teleported to the position of the logical ancilla as shown in~\autoref{fig:overview}{d}. 
Alternatively, the control patch can be measured such that one teleports the control to the ancilla as shown in~\autoref{fig:cnotmove_a}. 
To exploit a high level of flexibility of those teleportations, consider \textit{tree} structures as displayed in pink in~\autoref{fig:overview}{b}, with three terminal nodes, one for each of the control, target and ancilla patches.
Thus, the path between the control and target patch is extended by a branch to the position of the ancilla patch.
On the microscopic level, these tree structures are well-defined as elaborated in the following example.
\begin{example}         Consider~\autoref{fig:cnotmove_b} as example for a CNOT + teleportation step laid out on an architecture with distance $d=7$ color codes. %
    Assume that one wants to move the control qubit to the position of the blue patch, thus this patch is chosen as logical ancilla for the CNOT + teleportation scheme. 
    By performing an $X_LX_L$ measurement along the route indicated by the violet dashed line, followed by a $Z_LZ_L$ measurement along the orange dashed line, as well as the corrections as displayed in~\autoref{fig:cnotmove_a}, one can achieve the desired teleportation-based movement 
    of the control qubit, while executing the logical CNOT gate.%
\end{example}
Note that two dotted lines are ``entering'' the blue ancilla patch. 
This is only possible because the color code hosts both an $X_L$ and $Z_L$ operator along each boundary. 
Due to this freedom one could place the logical ancilla on any free spot that allows for a proper connection between the data patches. 
In case of the surface code, the choice of logical ancilla positions would be more restricted due to each boundary supporting only one logical Pauli, either $X_L$ or $Z_L$, but not both.

The adaption of the measurement-based scheme for the CNOT gate can be proven straightforwardly with the ZX-calculus~\cite{wetering_zx-calculus_2020, KissingerWetering2024Book} as shown in~\autoref{fig:zxderivation} for the CNOT + teleportation of control to ancilla position. 
The derivation works analogously for teleporting the target to the ancilla position.

\subsection{Exploiting Movable Logical Qubits at the Macroscopic Level}\label{sec-mainidea-macro}

\begin{figure}[t]
    \centering
    \includegraphics[width = 0.6\linewidth]{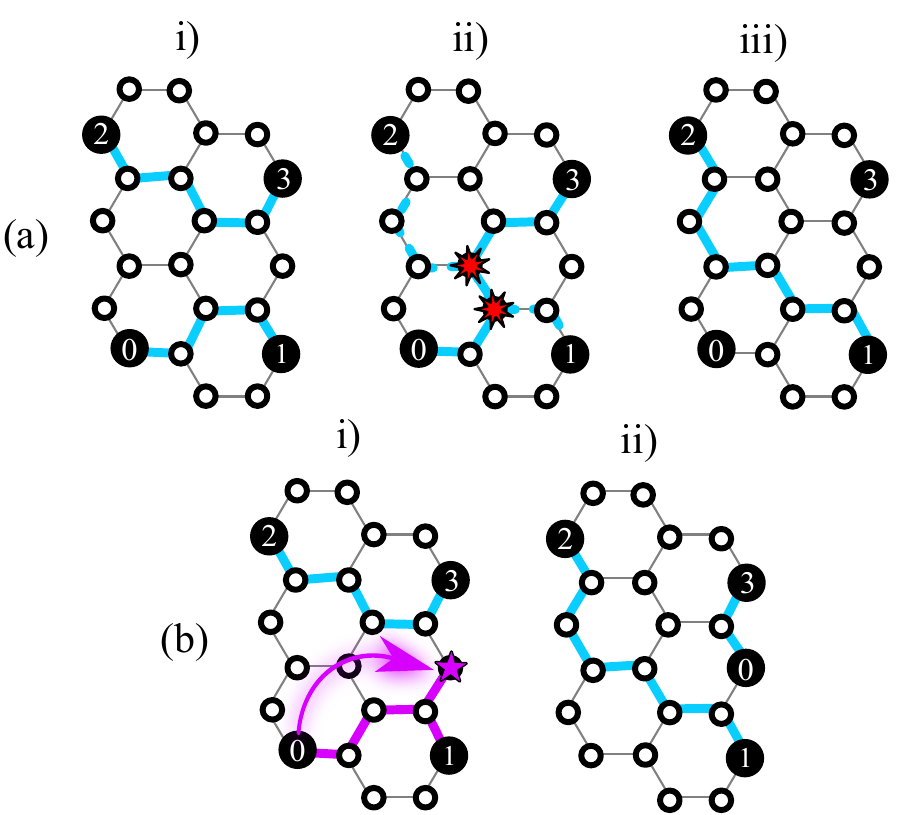}
    \caption{CNOT + teleportation at the macroscopic level. (a) Routing of the given circuit with four CNOT gates and two logical layers leads to three routed layers if the choice of qubit mapping is static. (b) With a suitable choice of a CNOT + teleportation one can route the logical circuit in two routed layers.}
    \label{fig:example_movement}
\end{figure}
Utilizing teleportation-based movements as introduced above enables us to go beyond the place-and-route paradigm on the macroscopic level of compilation. 
One can abstract away from the specific structure of the color code patches and simplify the picture to a macroscopic (hexagonal) routing graph $\mathcal{R}$ whose vertices represent color code patches. 
How the overall routed depth can be reduced by exploiting teleportation-based movements is explained by the following example.
\begin{example}

    Consider a logical circuit comprising the first logical layer CNOT(2,3) and CNOT(0,1), and the second layer CNOT(0,3) and CNOT(2,1).
    Its logical depth is $d_\mathrm{L}=2$.
    Given a specific, fixed logical data qubit mapping as displayed in ~\autoref{fig:example_movement}{a}, the first two gates can be routed simultaneously in layer (i). 
    In layer (ii), one can route CNOT(0,3) but the attempt of routing CNOT(2,1) in parallel inevitably fails due to a crossing of paths such that the latter gate must be relegated to layer (iii). 
    Thus, the routed depth is $\widetilde{d}=3$. 
    In contrast to this, starting with the same fixed mapping in (b) but extending one of the gates in layer (i) to a CNOT + teleportation step (pink tree), it is possible to route the remaining two gates simultaneously in layer (ii), since crossings could be avoided. 
    By exploiting movable logical qubits, the routed depth could be reduced to the logical depth of the logical circuit, i.e., $\widetilde{d}=d_\mathrm{L}=2$.

\end{example}

To illustrate how to use this degree of freedom in practice let us return to ~\autoref{fig:overview}. 
Assume a given macroscopic routing graph $\mathcal{R}$ as displayed in ~\autoref{fig:overview}{b}, and four logical gates in a logical/routed layer: CNOT(1,12), CNOT(2,10), CNOT(11,17), and CNOT(14,3). 
All these gates can be routed without overlaps. %
Beyond the standard CNOT gates in~\autoref{fig:overview}{b}, CNOT combined with teleportation (\autoref{fig:overview}{c}) enables us to reduce the routed depth in subsequent logical layers.
Given the routing of the current layer as displayed in
a 1D linearized view in ~\autoref{fig:overview}{e} one can search for beneficial tree structures (indicated in pink) based on the routings of a finite number $k$ of subsequent logical layers in the circuit as indicated in ~\autoref{fig:overview}{f}.
For example, the teleportation based on the tree structure in ~\autoref{fig:overview}{b} reduces the routed depth $\widetilde{\ell}$ of $k$ future logical layers from ~\autoref{fig:overview}{f}. 
This heuristic procedure is detailed in the next section.

\section{Proposed Heuristic Approach}\label{sec-heuristic}

\begin{figure}[t]
    \centering
    \includegraphics[width = 0.65\linewidth]{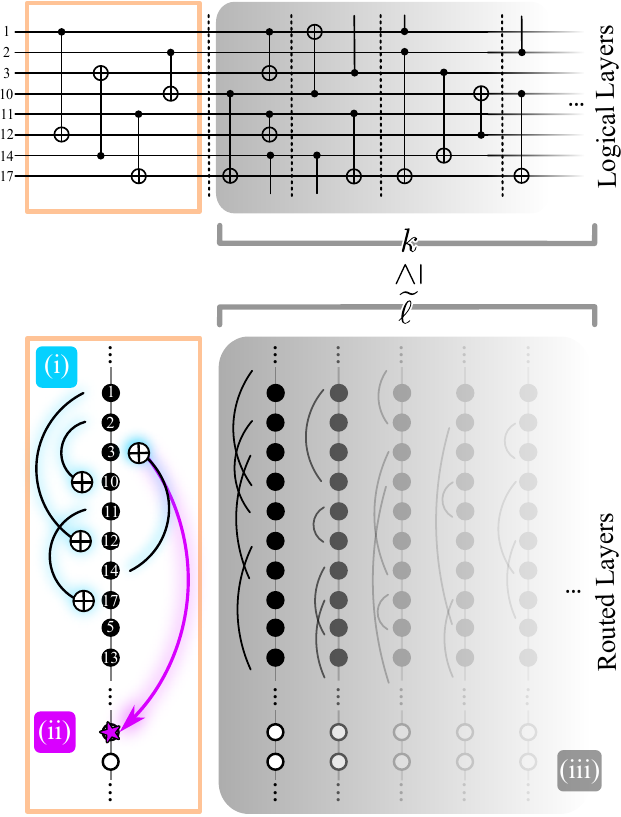}

    \caption{
    Heuristic optimization of routed depth with CNOT + teleportation.
    Top: The logical circuit is split in logical layers (separated by dashed lines), with the first enclosed by an orange box.
    Bottom: Logical circuit compilation into routed layers.
    First, the CNOTs of the first logical layer are routed in the standard way as indicated in blue (i). 
    Then, based on the next $k$ logical layers, tree structures (pink) are searched that minimize the number of routed layers $\widetilde{\ell}$ (ii).
    Afterwards (iii), the $\widetilde{\ell}$ resulting routed layers are fixed in the schedule.
    The procedure is repeated with the next logical layer (i).
    }
    \label{fig:sliding-window}
\end{figure}
The overall aim is to route the gates of a given circuit with the lowest routed depth $\widetilde{d}$. 
To this end, it is useful to exploit movable logical qubits during the execution of CNOT gates, as detailed in the previous section.
We now explain the heuristic ``sliding-window'' strategy with lookahead in ~\autoref{sec-sliding-window}, followed by an explanation of a simulated annealing based optimization subroutine in ~\autoref{sec-sa}, and the underlying shortest-first routing routine in ~\autoref{sec-vdp-subroutine}.

\subsection{Sliding Window Approach}\label{sec-sliding-window}

The algorithm used in this work is summarized in ~\autoref{fig:sliding-window}. 
(i) Starting with the first logical layer $j=0$ of the circuit (top orange box), one routes as many gates as possible by a shortest-first routing (\autoref{sec-vdp-subroutine}) as indicated by blue shaded gates (bottom orange box). 
This constitutes the first routed layer.
Gates that cannot be routed anymore because of path overlaps are pushed to the next logical layer.
The logical structure of the upcoming layers is updated correspondingly. 
(ii) Having fixed the routes for the first routed layer, one looks for extensions of the paths to trees and corresponding teleportations of the control-to-ancilla or target-to-ancilla position, based on the next $k$ logical lookahead layers of the circuit. 
If an optimized choice of tree extensions reduces the number of routed layers $\widetilde{\ell} \geq k$, the teleportations are applied. 
(iii) Finally, one fixes the corresponding $\widetilde{\ell}$ layers in the schedule. 
Then, the next logical layer $j'=j+k+1$ is considered and the procedure is repeated until all gates of the logical circuit are routed.
Note that if, in step (ii), no trees are found, one directly jumps to the next logical layer $j'=j+1$ and repeats the procedure with step (i).

Teleportations optimized for the next $k$ logical lookahead layers may temporarily create configurations in which certain logical data qubits become effectively disconnected from the routing space, as they are fully enclosed by other logical data patches. 
To permit such configurations---temporarily advantageous but harmful in the long run---idle data qubits are teleported during step (i) via standard lattice surgery teleportation into ``gaps'' generated when logical data qubits were moved in previous iterations. 
In this way, a large variety of CNOT + teleportation steps can be exploited while preserving a consistent layout structure in the long term, such as the pair layout shown in ~\autoref{fig:overview}{b}.

\subsection{Simulated Annealing for Tree Search}\label{sec-sa}

The search for trees in step (ii) is performed by applying simulated annealing~\cite{kirkpatrick_optimization_1983}. 
The inputs for this routine are the current routed layer (routes of blue gates in (i)) and the $k$ subsequent logical lookahead layers.
A sample solution is initialized by extending the paths of the blue gates to random trees on the ancilla space. 
In addition, note that it is in principle also possible to place the ancilla along the path without explicitly extending the path to a tree and teleporting towards a patch \textit{on} the path. 
The objective function to minimize is the number of routed layers $\widetilde{\ell}$ of the next $k$ logical lookahead layers, which is computed explicitly with the shortest-first routing (\autoref{sec-vdp-subroutine}).
Furthermore, a simulated annealing procedure requires the definition of a ``neighborhood'' to choose a new sample solution from.
The neighborhood of a tree is defined by all variations of the tree in which the ancilla patch positions lie within a radius $r$
of edges from the initial ancilla patch position on $\mathcal{R}$. One randomly teleports the control or target qubit to the new ancilla position.
The simulated annealing procedure therefore searches tree structures based on the routed depth $\widetilde{\ell}$ of the next $k$ logical lookahead layers and attempts to find solutions which reduce $\widetilde{\ell}$.

\subsection{Shortest-First Routing Subroutine}\label{sec-vdp-subroutine}

The routing of $k$ logical layers is performed in the same spirit as in~\cite{herzog_lattice_2025}. 
Given a circuit of CNOT gates only, one can formalize the problem as \textit{vertex disjoint path} (VDP) problem~\cite{bacco_shortest_2014, chuzhoy_approximating_2015, chuzhoy_improved_2016}. 
Based on a simple approximation scheme of the VDP problem~\cite{chuzhoy_approximating_2015}, routing $k$ layers of CNOT gates means, that given a logical layer, one begins by computing the shortest paths for all CNOT gates using Dijkstra’s algorithm. 
From those, one selects the shortest path and fixes it in the current routed layer and removes the used nodes from~$\mathcal{R}$. 
The procedure is repeated until no further paths fit on $\mathcal{R}$. 
All routes found in this procedure constitute a routed layer. 
If not all CNOT gates of the given logical layer could be routed, they are pushed in the next logical layer and all logical layers are updated correspondingly in the DAG. 
Then, the procedure is repeated for the next logical layer until all gates in the initially $k$ logical layers are routed and turned into $\widetilde{\ell}$ routed layers.

Note that this procedure can be greedily generalized to include T gates %
with a factory reset time $t$~\cite{herzog_lattice_2025}. 
In the same spirit, the proposed optimization with CNOT + teleportation steps can be generalized to include T gates as well. 
Assume a T state injection scheme using a CNOT gate; this CNOT gate can also be used for a CNOT + teleportation step. 
The proposed optimization scheme with CNOT and T gates is available on GitHub \url{https://github.com/munich-quantum-toolkit/qecc}.

\section{Performance Evaluation}\label{sec-numerics}

\begin{figure}[b]
    \centering
    \includegraphics[width=0.49\linewidth]{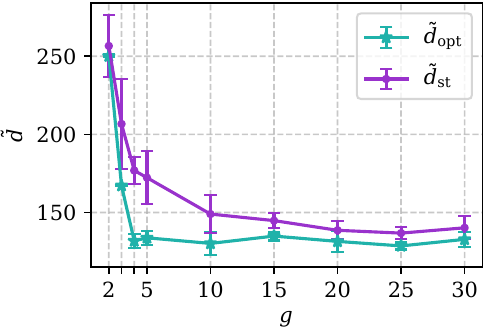}
    \includegraphics[width=0.49\linewidth]{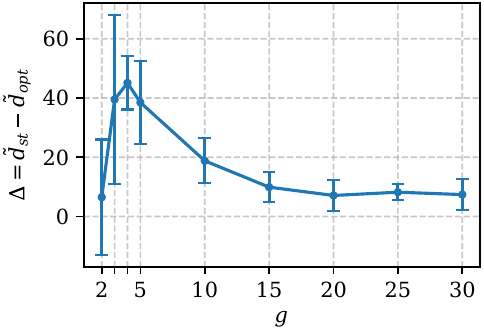}
    \caption{$\widetilde{d}$ (left) and $\Delta=\widetilde{d}_{\mathrm{st}} - \widetilde{d}_{\mathrm{opt}}$ (right) under variation of $g$ with a fixed number of gates $G=500$ and $q=60$ logical data qubits in a triple layout. 
    For each data point $10$ random circuits were sampled.}
    \label{fig:varj}
\end{figure}

\begin{table*}[ht]
\centering
\scriptsize
\caption{Comparison of scheduling performance across different layouts for $q=120$ logical qubits. For each layout the smallest unit of data patches is displayed together with the asymptotic layout density $c$ which is the ratio between the number of data patches and the total number of patches. A randomly sampled circuit type with $g=8$ gates per logical layer was used. $\widetilde{d}_{\mathrm{st}}$ and $\widetilde{d}_{\mathrm{opt}}$ denote the routed depths of the standard approach and the proposed optimized approach using CNOT + teleportation steps, with both absolute values and a linear fit; $d_\mathrm{L}$ is the logical circuit depth.
Values represent mean $\pm$ standard deviation. For each entry $10$ random circuits were sampled.
}
\begin{tabular}{cccccccccc}
\toprule
{Layout} & 
Logical Depth ${d_\mathrm{L}}$ & 
Standard Method ${\widetilde{d}_{\mathrm{st}}}$ & 
Proposed Method ${\widetilde{d}_{\mathrm{opt}}}$ & 
${\widetilde{r} = \frac{\widetilde{d}_{\mathrm{opt}} - d_\mathrm{L}}{\widetilde{d}_{\mathrm{st}} - d_\mathrm{L}}}$ & 
${\Delta = \widetilde{d}_{\mathrm{st}} - \widetilde{d}_{\mathrm{opt}}}$ & $\frac{\Delta}{d_{\mathrm{st}}}$\\
\midrule
\multirow{4}{*}{\begin{minipage}{0.05\textwidth}
    \centering
    Single $c=\frac{1}{8}$\\[2pt]
    \includegraphics[width=0.75\linewidth]{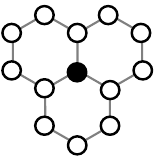}
\end{minipage}} 
& $40$ & $47.9 \pm 2.7$ & $40.2 \pm 0.4$ & $3.8 \pm 8.0$\% & $7.7 \pm 2.8$ & $15.8 \pm 5.1$\% \\
& $80$ & $95.4 \pm 3.7$ & $80.3 \pm 0.5$ & $2.4 \pm 3.8$\% & $15.1 \pm 4.0$ & $15.7 \pm 3.6$\% \\
& $160$ & $192.5 \pm 6.8$ & $160.4 \pm 0.7$ & $1.5 \pm 2.3$\% & $32.1 \pm 7.1$ & $16.6 \pm 3.1$\% \\
& $320$ & $384.5 \pm 13.1$ & $320.7 \pm 0.8$ & $1.2 \pm 1.3$\% & $63.8 \pm 13.3$ & $16.5 \pm 2.9$\% \\
     & Fit: &  $\widetilde{d}_{\mathrm{st}} = 1.2\, d_\mathrm{L} -0.4 $& $\widetilde{d}_{\mathrm{opt}} = 1.0\, d_\mathrm{L} +0.1  $ & & \\
\midrule
\multirow{4}{*}{\begin{minipage}{0.06\textwidth}
    \centering
    Pair $c=\frac{1}{4}$\\[2pt]
    \includegraphics[width=0.85\linewidth]{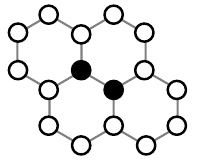}
\end{minipage}} 
& $40$ & $60.5 \pm 6.2$ & $48.3 \pm 1.4$ & $43.5 \pm 12.3$\% & $12.2 \pm 5.9$ & $19.4 \pm 7.5$\% \\
& $80$ & $119.5 \pm 9.3$ & $97.5 \pm 2.3$ & $46.2 \pm 11.0$\% & $22.0 \pm 8.8$ & $18.0 \pm 5.8$\% \\
& $160$ & $237.6 \pm 16.0$ & $193.9 \pm 3.9$ & $45.0 \pm 8.5$\% & $43.7 \pm 14.2$ & $18.1 \pm 4.8$\% \\
& $320$ & $473.6 \pm 31.0$ & $385.6 \pm 7.1$ & $43.9 \pm 7.6$\% & $88.0 \pm 27.2$ & $18.3 \pm 4.5$\% \\
   & Fit: &  $\widetilde{d}_{\mathrm{st}} = 1.5\, d_\mathrm{L} +1.5$& $\widetilde{d}_{\mathrm{opt}} = 1.2\, d_\mathrm{L}+0.8 $ & & \\
\midrule
\multirow{4}{*}{\begin{minipage}{0.07\textwidth}
    \centering
    Triple $c=\frac{3}{10}$\\[2pt]
    \includegraphics[width=0.8\linewidth]{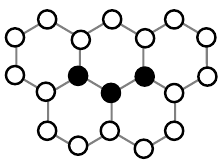}
\end{minipage}} 
& $40$ & $78.3 \pm 9.6$ & $60.1 \pm 3.4$ & $54.9 \pm 11.6$\% & $18.2 \pm 8.1$ & $22.3 \pm 8.0$\% \\
& $80$ & $156.5 \pm 17.4$ & $120.3 \pm 5.2$ & $54.4 \pm 9.1$\% & $36.2 \pm 14.8$ & $22.4 \pm 6.8$\% \\
& $160$ & $312.6 \pm 30.5$ & $238.8 \pm 8.0$ & $53.2 \pm 8.8$\% & $73.8 \pm 26.4$ & $23.0 \pm 6.3$\% \\
& $320$ & $625.1 \pm 56.6$ & $480.0 \pm 15.7$ & $53.7 \pm 7.6$\% & $145.1 \pm 47.7$ & $22.7 \pm 5.6$\% \\
  & Fit: &  $\widetilde{d}_{\mathrm{st}} = 2.0\, d_\mathrm{L}+0.2$ & $\widetilde{d}_{\mathrm{opt}} = 1.5\, d_\mathrm{L}-0.1$ & & \\
\midrule
\multirow{4}{*}{\begin{minipage}{0.06\textwidth}
    \centering
    Hex $c=\frac{3}{7}$\\[2pt]
    \includegraphics[width=0.8\linewidth]{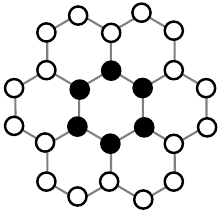}
\end{minipage}} 
& $40$ & $90.9 \pm 4.6$ & $75.9 \pm 3.0$ & $70.9 \pm 6.7$\% & $15.0 \pm 4.4$ & $16.4 \pm 4.2$\% \\
& $80$ & $181.0 \pm 8.6$ & $151.0 \pm 4.7$ & $70.7 \pm 6.1$\% & $30.0 \pm 8.4$ & $16.4 \pm 3.9$\% \\
& $160$ & $365.3 \pm 18.1$ & $304.4 \pm 10.4$ & $70.6 \pm 5.3$\% & $60.9 \pm 15.3$ & $16.5 \pm 3.4$\% \\
& $320$ & $731.0 \pm 40.4$ & $609.7 \pm 18.0$ & $70.9 \pm 5.1$\% & $121.3 \pm 31.7$ & $16.4 \pm 3.5$\% \\
 & Fit: &  $\widetilde{d}_{\mathrm{st}} =2.3 \, d_\mathrm{L} - 1.2$& $\widetilde{d}_{\mathrm{opt}} =1.9 \, d_\mathrm{L}-1.0$ & & \\
\bottomrule
\label{tab:comparison_q120}
\end{tabular}
\end{table*}

This section evaluates the proposed compilation method with movable logical qubits and identifies regimes where it performs best.

For this comparison, we consider the application of the shortest-first routing described in ~\autoref{sec-vdp-subroutine} for the whole logical circuit as a standard routing baseline with static logical data patch locations, leading to routed depth $\widetilde{d}_{\mathrm{st}}$. 
This is compared to the proposed approach with teleportation-based movements. %
We denote the resulting routed depth as $\widetilde{d}_{\mathrm{opt}}$. 
The following proof-of-principle comparison is performed using random initial mappings and random CNOT circuits with varying numbers of gates per logical layer.
Random initial mappings of data qubit labels are used to isolate the effect of the proposed method.
Note that is reasonable to assume that initial mapping optimization~\cite{molavi_dependency-aware_2024} is assumed to offer limited benefit for growing circuit depth.

The random circuits used here are constructed in such a way that they can vary in their ratio between number of logical CNOT gates $G$ and their depth $d_\mathrm{L}$. 
This means that one can choose a number $g$ of randomly sampled CNOT gates per layer in the DAG such that one has in total $G = g\cdot  d_{\mathrm{L}}$ gates in the circuit. %

The interplay between the density of the circuit (determined \mbox{by $g$}) and the density of the layout play a major role how the proposed method performs. 
The layout describes the ratio $c$ between the data and total patches on the macroscopic routing graph and how they are placed, as for instance the pair layout 
in~\autoref{fig:overview}{b}. 
Further layouts are displayed in the left-most column of ~\autoref{tab:comparison_q120}.

To study this interplay, let us fix the layout and the number of logical data qubits first and vary the value $g$ with a fixed total number of gates $G$. For the optimization $k=5$ logical lookahead layers were chosen with radius $r=10$. 
In ~\autoref{fig:varj} one can see that the optimization performs best with rather low choices of $g$. 
If the circuit is too dense, then there are already too many paths on the routing graph once one attempts to find tree extensions for optimizations such that the benefit saturates.
On the other hand, if the density of the circuit is too low, there is not enough parallelism to exploit such that the optimizations become negligible or unreliable.%

This study allows us to fix $g$ in a suitable regime for the proposed method. 
Varying the logical depth $d_\mathrm{L}$ and the layouts provides further information about the aforementioned interplay as displayed in ~\autoref{tab:comparison_q120}. 
With $q=120$ logical data qubits in total and $g=8$, both the layouts and the logical depth were varied. One can observe that the relative layer overhead $\widetilde{r}$ is very low for the single layout, which is the sparsest layout.
In other words, the proposed method can yield routed depths very close to the logical depth, which is the lower bound on the routed depth. With increasing layout density, also $\widetilde{r}$ grows, since parallel routing and tree search become more challenging. Without taking the relation to the logical depth into consideration, the relative layer reductions $\Delta / d_\mathrm{st} = (\widetilde{d}_{\mathrm{st}} - \widetilde{d}_{\mathrm{opt}})/d_\mathrm{st}$ are the highest for layouts with intermediate density, such as the triple layout. %

Irrespective of the specific layout, the absolute difference $\Delta = \widetilde{d}_{\mathrm{st}} - \widetilde{d}_{\mathrm{opt}}$ increases with growing logical depth for any layout. 
This illustrates that consecutive teleportation-based movements of data patches can indeed lead to consecutive improvements throughout the whole circuit.
This crucial observation is also underpinned by the lower slopes in the linear fits for $\widetilde{d}_\mathrm{opt}$ than for~$\widetilde{d}_{\mathrm{st}}$.

Overall, our results demonstrate that exploiting teleportation-based movements during lattice surgery compilation can yield significant reductions in routed depths compared to a standard routing with fixed data qubit positions.

\section{Conclusion}\label{sec-conclusion}
Our work constitutes the first attempt to go beyond the place-and-route paradigm for quantum hardware architectures with static physical qubits and with focus on reducing schedule depth. 
We provided a proof-of-principle software implementation that exploits the inherent degree of freedom of teleporting the control or target logical during the logical lattice surgery CNOT gate. 
Furthermore, we explored the relevant trade-offs in the density of the layout and the logical circuit, identifying regimes in which significant reductions can be achieved.

Going forward, it will be useful to extend the evaluation of our software implementation to studies on specific quantum algorithms, or to adapt it to multi-qubit Pauli measurements~\cite{litinski_game_2019}.
Finally, as a promising direction for future work we envision that the proposed methods can be used to tackle growing schedule depths if defects are considered---thereby acting as the logical counterpart to existing proposals on the physical level~\cite{debroy_luci_2024,leroux_snakes_2024}.

    \begin{acks}
We thank Tom Peham for fruitful exchange regarding ZX calculus. L.H., L.B. and R.W. acknowledge funding from the European Research Council (ERC) under the European Union’s Horizon 2020 research and innovation program (grant agreement No.\ 101001318) and Millenion (grant agreement No.\ 101114305). 
This work was part of the Munich Quantum Valley, which is supported by the Bavarian state government with funds from the Hightech Agenda Bayern Plus.
This work was funded by the Deutsche Forschungsgemeinschaft (DFG, German Research Foundation, No. 563402549).
A.K. acknowledges support from the NSF (QLCI, Award No. OMA-2120757), IARPA and the Army Research Office (ELQ Program, Cooperative Agreement No. W911NF-23-2-0219).
\end{acks}

\clearpage
\balance
\printbibliography

@String{Computing = "Computing" }

@String{Computer = "{IEEE} Computer" }

@misc{fowler2019lowoverheadquantumcomputation,
      title={Low overhead quantum computation using lattice surgery}, 
      author={Austin G. Fowler and Craig Gidney},
      year={2019},
      eprint={1808.06709},
      archivePrefix={arXiv},
      primaryClass={quant-ph},
      url={https://arxiv.org/abs/1808.06709}, 
}

@misc{sharma_space-time_2025,
	title = {Space-Time Optimisations for Early Fault-Tolerant Quantum Computation},
	url = {http://arxiv.org/abs/2511.08848},
	doi = {10.48550/arXiv.2511.08848},
	number = {{arXiv}:2511.08848},
	publisher = {{arXiv}},
	author = {Sharma, Sanaa and Murali, Prakash},
	urldate = {2025-11-13},
	date = {2025-11-13},
	eprinttype = {arxiv},
	eprint = {2511.08848 [quant-ph]},
}

@misc{mcardle2025fastcuriousacceleratefaulttolerant,
      title={The Fast for the Curious: How to accelerate fault-tolerant quantum applications}, 
      author={Sam McArdle and Alexander M. Dalzell and Aleksander Kubica and Fernando G. S. L. Brandão},
      year={2025},
      eprint={2510.26078},
      archivePrefix={arXiv},
      primaryClass={quant-ph},
      url={https://arxiv.org/abs/2510.26078}, 
}

@inproceedings{childs-dag,
  doi = {10.4230/LIPICS.TQC.2019.3},
  
  url = {https://drops.dagstuhl.de/entities/document/10.4230/LIPIcs.TQC.2019.3},
  
  author = {Childs, Andrew M. and Schoute, Eddie and Unsal, Cem M.},
  
  language = {en},
  
  title = {Circuit Transformations for Quantum Architectures},
  
  publisher = {Schloss Dagstuhl – Leibniz-Zentrum für Informatik},
  
  year = {2019},
  
  copyright = {Creative Commons Attribution 3.0 Unported license}
}

@misc{javadiabhari2024quantumcomputingqiskit,
      title={Quantum computing with Qiskit}, 
      author={Ali Javadi-Abhari and Matthew Treinish and Kevin Krsulich and Christopher J. Wood and Jake Lishman and Julien Gacon and Simon Martiel and Paul D. Nation and Lev S. Bishop and Andrew W. Cross and Blake R. Johnson and Jay M. Gambetta},
      year={2024},
      eprint={2405.08810},
      archivePrefix={arXiv},
      primaryClass={quant-ph},
      url={https://arxiv.org/abs/2405.08810}, 
}

@article{kirkpatrick_optimization_1983,
	title = {Optimization by Simulated Annealing},
	volume = {220},
	issn = {0036-8075, 1095-9203},
	url = {https://www.science.org/doi/10.1126/science.220.4598.671},
	doi = {10.1126/science.220.4598.671},
	pages = {671--680},
	number = {4598},
	journaltitle = {Science},
	author = {Kirkpatrick, S. and Gelatt, C. D. and Vecchi, M. P.},
	urldate = {2025-11-04},
	date = {1983-05-13},
	langid = {english},
}

@misc{cross_improved_2025,
	title = {Improved {QLDPC} Surgery: Logical Measurements and Bridging Codes},
	url = {http://arxiv.org/abs/2407.18393},
	doi = {10.48550/arXiv.2407.18393},
	shorttitle = {Improved {QLDPC} Surgery},
	number = {{arXiv}:2407.18393},
	publisher = {{arXiv}},
	author = {Cross, Andrew W. and He, Zhiyang and Rall, Patrick J. and Yoder, Theodore J.},
	urldate = {2025-11-04},
	date = {2025-10-27},
	eprinttype = {arxiv},
	eprint = {2407.18393 [quant-ph]},
}

@misc{baspin_fast_2025,
	title = {Fast surgery for quantum {LDPC} codes},
	url = {http://arxiv.org/abs/2510.04521},
	doi = {10.48550/arXiv.2510.04521},
	number = {{arXiv}:2510.04521},
	publisher = {{arXiv}},
	author = {Baspin, Nouédyn and Berent, Lucas and Cohen, Lawrence Z.},
	urldate = {2025-11-04},
	date = {2025-10-06},
	eprinttype = {arxiv},
	eprint = {2510.04521 [quant-ph]},
}

@misc{cowtan_ssip_2024,
	title = {{SSIP}: automated surgery with quantum {LDPC} codes},
	url = {http://arxiv.org/abs/2407.09423},
	doi = {10.48550/arXiv.2407.09423},
	shorttitle = {{SSIP}},
	number = {{arXiv}:2407.09423},
	publisher = {{arXiv}},
	author = {Cowtan, Alexander},
	urldate = {2025-11-04},
	date = {2024-07-12},
	eprinttype = {arxiv},
	eprint = {2407.09423 [quant-ph]},
}

@misc{eisert_mind_2025,
	title = {Mind the gaps: The fraught road to quantum advantage},
	url = {http://arxiv.org/abs/2510.19928},
	doi = {10.48550/arXiv.2510.19928},
	shorttitle = {Mind the gaps},
	number = {{arXiv}:2510.19928},
	publisher = {{arXiv}},
	author = {Eisert, Jens and Preskill, John},
	urldate = {2025-10-30},
	date = {2025-10-22},
	eprinttype = {arxiv},
	eprint = {2510.19928 [quant-ph]},
}

@book{KissingerWetering2024Book,
    author = {Kissinger, Aleks and van de Wetering, John},
    title = {{Picturing Quantum Software: An Introduction to the ZX-Calculus and Quantum Compilation}},
    year = {2024},
    publisher = {Preprint},
}

@misc{wetering_zx-calculus_2020,
	title = {{ZX}-calculus for the working quantum computer scientist},
	url = {http://arxiv.org/abs/2012.13966},
	doi = {10.48550/arXiv.2012.13966},
	number = {{arXiv}:2012.13966},
	publisher = {{arXiv}},
	author = {Wetering, John van de},
	urldate = {2025-10-30},
	date = {2020-12-27},
	eprinttype = {arxiv},
	eprint = {2012.13966 [quant-ph]},
}

@misc{robertson_resource_2025,
	title = {A Resource Allocating Compiler for Lattice Surgery},
	url = {http://arxiv.org/abs/2506.04620},
	doi = {10.48550/arXiv.2506.04620},
	number = {{arXiv}:2506.04620},
	publisher = {{arXiv}},
	author = {Robertson, Alan and Gao, Haowen and Sanders, Yuval R.},
	urldate = {2025-10-27},
	date = {2025-06-05},
	eprinttype = {arxiv},
	eprint = {2506.04620 [quant-ph]},
}

@misc{herzog_lattice_2025,
	title = {Lattice Surgery Compilation Beyond the Surface Code},
	url = {http://arxiv.org/abs/2504.10591},
	doi = {10.48550/arXiv.2504.10591},
	number = {{arXiv}:2504.10591},
	publisher = {{arXiv}},
	author = {Herzog, Laura S. and Berent, Lucas and Kubica, Aleksander and Wille, Robert},
	urldate = {2025-10-10},
	date = {2025-08-01},
	eprinttype = {arxiv},
	eprint = {2504.10591 [quant-ph]},
}

@misc{koutsioumpas_colour_2025,
	title = {Colour Codes Reach Surface Code Performance using Vibe Decoding},
	url = {http://arxiv.org/abs/2508.15743},
	doi = {10.48550/arXiv.2508.15743},
	abstract = {Two-dimensional quantum colour codes hold significant promise for quantum error correction, offering advantages such as planar connectivity and low overhead logical gates. Despite their theoretical appeal, the practical deployment of these codes faces challenges due to complex decoding requirements compared to surface codes. This paper introduces vibe decoding which, for the first time, brings colour code performance on par with the surface code under practical decoding. Our approach leverages an ensemble of belief propagation decoders - each executing a distinct serial message passing schedule - combined with localised statistics post-processing. We refer to this combined protocol as {VibeLSD}. The {VibeLSD} decoder is highly versatile: our numerical results show it outperforms all practical existing colour code decoders across various syndrome extraction schemes, noise models, and error rates. By estimating qubit footprints through quantum memory simulations, we show that colour codes can operate with overhead that is comparable to, and in some cases lower than, that of the surface code. This, combined with the fact that localised statistics decoding is a parallel algorithm, makes {VibeLSD} suitable for implementation on specialised hardware for real-time decoding. Our results establish the colour code as a practical architecture for near-term quantum hardware, providing improved compilation efficiency for both Clifford and non-Clifford gates without incurring additional qubit overhead relative to the surface code.},
	number = {{arXiv}:2508.15743},
	publisher = {{arXiv}},
	author = {Koutsioumpas, Stergios and Noszko, Tamas and Sayginel, Hasan and Webster, Mark and Roffe, Joschka},
	%urldate = {2025-08-22},
	date = {2025-08-21},
	eprinttype = {arxiv},
	eprint = {2508.15743 [quant-ph]},
	keywords = {Quantum Physics, Computer Science - Information Theory, Mathematics - Information Theory},
	file = {Preprint PDF:C\:\\Users\\Laura\\Zotero\\storage\\SZ7B99FN\\Koutsioumpas et al. - 2025 - Colour Codes Reach Surface Code Performance using Vibe Decoding.pdf:application/pdf;Snapshot:C\:\\Users\\Laura\\Zotero\\storage\\22XHVJEF\\2508.html:text/html},
}

@misc{leblond2025quantumresourcecomparisonleading,
      title={Quantum Resource Comparison for Two Leading Surface Code Lattice Surgery Approaches}, 
      author={Tyler LeBlond and Ryan S. Bennink},
      year={2025},
      eprint={2506.08182},
      archivePrefix={arXiv},
      primaryClass={quant-ph},
      url={https://arxiv.org/abs/2506.08182}, 
}

@article{Kubica2023,
  title = {Efficient color code decoders in $d\geq 2$ dimensions from toric code decoders},
  volume = {7},
  ISSN = {2521-327X},
  url = {http://dx.doi.org/10.22331/q-2023-02-21-929},
  DOI = {10.22331/q-2023-02-21-929},
  journal = {Quantum},
  publisher = {Verein zur Forderung des Open Access Publizierens in den Quantenwissenschaften},
  author = {Kubica,  Aleksander and Delfosse,  Nicolas},
  year = {2023},
  month = feb,
  pages = {929}
}

@misc{dalzell2023quantum,
      title={Quantum algorithms: A survey of applications and end-to-end complexities}, 
      author={Alexander M. Dalzell and Sam McArdle and Mario Berta and Przemyslaw Bienias and Chi-Fang Chen and András Gilyén and Connor T. Hann and Michael J. Kastoryano and Emil T. Khabiboulline and Aleksander Kubica and Grant Salton and Samson Wang and Fernando G. S. L. Brandão},
      year={2023},
      eprint={2310.03011},
      archivePrefix={arXiv},
      primaryClass={quant-ph},
      url={https://arxiv.org/abs/2310.03011}, 
}

@article{Shor1995,
  title = {Scheme for reducing decoherence in quantum computer memory},
  volume = {52},
  ISSN = {1094-1622},
  url = {http://dx.doi.org/10.1103/PhysRevA.52.R2493},
  DOI = {10.1103/physreva.52.r2493},
  number = {4},
  journal = {Physical Review A},
  publisher = {American Physical Society (APS)},
  author = {Shor,  Peter W.},
  year = {1995},
  month = oct,
  pages = {R2493–R2496}
}

@article{kitaev1997quantum,
  title={Quantum computations: algorithms and error correction},
  author={Kitaev, A Yu},
  journal={Russian Mathematical Surveys},
  volume={52},
  number={6},
  pages={1191},
  year={1997},
  publisher={IOP Publishing}
}

@article{Dennis_2002,
	title        = {Topological quantum memory},
	author       = {Dennis, Eric and Kitaev, Alexei and Landahl, Andrew and Preskill, John},
	year         = 2002,
	journal      = {Journal of Mathematical Physics},
	publisher    = {AIP Publishing},
	volume       = 43,
	number       = 9,
	pages        = {4452–4505},
	doi          = {10.1063/1.1499754}
}

@article{Kitaev_2003,
	title        = {Fault-tolerant quantum computation by anyons},
	author       = {Kitaev, A.Yu.},
	year         = 2003,
	journal      = {Annals of Physics},
	publisher    = {Elsevier BV},
	volume       = 303,
	number       = 1,
	pages        = {2–30},
	doi          = {10.1016/s0003-4916(02)00018-0}
}

@phdthesis{kubica_abc_2018,
	title        = {The ABCs of the Color Code: A Study of Topological Quantum Codes as Toy Models for Fault-Tolerant Quantum Computation and Quantum Phases Of Matter},
	author       = {Kubica,  Aleksander},
	year         = 2018,
	publisher    = {California Institute of Technology},
	doi          = {10.7907/059V-MG69},
	copyright    = {No commercial reproduction,  distribution,  display or performance rights in this work are provided.},
}

@article{fowler_surface_2012,
	title        = {Surface codes: Towards practical large-scale quantum computation},
	shorttitle   = {Surface codes},
	author       = {Fowler, Austin G. and Mariantoni, Matteo and Martinis, John M. and Cleland, Andrew N.},
	volume       = 86,
	number       = 3,
	pages        = {032324},
	doi          = {10.1103/PhysRevA.86.032324},
	journaltitle = {Phys. Rev. A},
	date         = {2012-09-18},
	eprinttype   = {arxiv},
	eprint       = {1208.0928 [quant-ph]}
}

@inproceedings{chuzhoy_approximating_2015,
	title        = {On Approximating Node-Disjoint Paths in Grids},
	author       = {Chuzhoy, Julia and Kim, David H. K.},
	booktitle    = {Approximation, Randomization, and Combinatorial Optimization. Algorithms and Techniques ({APPROX}/{RANDOM} 2015)},
	publisher    = {Schloss Dagstuhl – Leibniz-Zentrum für Informatik},
	pages        = {187--211},
	doi          = {10.4230/LIPIcs.APPROX-RANDOM.2015.187},
	rights       = {https://creativecommons.org/licenses/by/3.0/legalcode},
	date         = 2015,
	langid       = {english}
}

@inproceedings{chuzhoy_improved_2016,
	title        = {Improved approximation for node-disjoint paths in planar graphs},
	author       = {Chuzhoy, Julia and Kim, David H. K. and Li, Shi},
	booktitle    = {Proceedings of the forty-eighth annual {ACM} symposium on Theory of Computing},
	publisher    = {Association for Computing Machinery},
	pages        = {556--569},
	doi          = {10.1145/2897518.2897538},
	date         = {2016-06-19}
}

@article{bacco_shortest_2014,
doi = {10.1088/1742-5468/2014/07/P07009},
url = {https://dx.doi.org/10.1088/1742-5468/2014/07/P07009},
year = {2014},
month = {jul},
publisher = {IOP Publishing and SISSA},
volume = {2014},
number = {7},
pages = {P07009},
author = {Bacco, C De and Franz, S and Saad, D and Yeung, C H},
title = {Shortest node-disjoint paths on random graphs},
journal = {Journal of Statistical Mechanics: Theory and Experiment}
}

@article{litinski_game_2019,
	title        = {A Game of Surface Codes: Large-Scale Quantum Computing with Lattice Surgery},
	shorttitle   = {A Game of Surface Codes},
	author       = {Litinski, Daniel},
	volume       = 3,
	pages        = 128,
	doi          = {10.22331/q-2019-03-05-128},
	journaltitle = {Quantum},
	date         = {2019-03-05},
	eprinttype   = {arxiv},
	eprint       = {1808.02892 [cond-mat, physics:quant-ph]}
}

@article{watkins_high_2024,
	title        = {A High Performance Compiler for Very Large Scale Surface Code Computations},
	author       = {Watkins, George and Nguyen, Hoang Minh and Watkins, Keelan and Pearce, Steven and Lau, Hoi-Kwan and Paler, Alexandru},
	volume       = 8,
	pages        = 1354,
	doi          = {10.22331/q-2024-05-22-1354},
	journaltitle = {Quantum},
	date         = {2024-05-22},
	eprinttype   = {arxiv},
	eprint       = {2302.02459 [quant-ph]}
}

@article{beverland_cost_2021,
	title        = {Cost of Universality: A Comparative Study of the Overhead of State Distillation and Code Switching with Color Codes},
	shorttitle   = {Cost of Universality},
	author       = {Beverland, Michael E. and Kubica, Aleksander and Svore, Krysta M.},
	volume       = 2,
	number       = 2,
	pages        = {020341},
	doi          = {10.1103/PRXQuantum.2.020341},
	journaltitle = {{PRX} Quantum}
}

@article{horsman_surface_2012,
	title        = {Surface code quantum computing by lattice surgery},
	author       = {Horsman, Dominic and Fowler, Austin G. and Devitt, Simon and Meter, Rodney Van},
	volume       = 14,
	number       = 12,
	pages        = 123011,
	doi          = {10.1088/1367-2630/14/12/123011},
	note         = {Publisher: {IOP} Publishing},
	journaltitle = {New J. Phys.},
	date         = {2012-12},
	langid       = {english}
}

@misc{bombin_introduction_2013,
	title        = {An Introduction to Topological Quantum Codes},
	author       = {Bombin, H.},
	publisher    = {{arXiv}},
	number       = {{arXiv}:1311.0277},
	doi          = {10.48550/arXiv.1311.0277},
	date         = {2013-11-01},
	eprinttype   = {arxiv},
	eprint       = {1311.0277 [quant-ph]}
}

@misc{landahl_quantum_2014,
      title={Quantum computing by color-code lattice surgery}, 
      author={Andrew J. Landahl and Ciaran Ryan-Anderson},
      year={2014},
      eprint={1407.5103},
      archivePrefix={arXiv},
      primaryClass={quant-ph},
}

@article{lao_mapping_2018,
	title        = {Mapping of lattice surgery-based quantum circuits on surface code architectures},
	author       = {Lao, L. and Wee, B. van and Ashraf, I. and Someren, J. van and Khammassi, N. and Bertels, K. and Almudever, C. G.},
	volume       = 4,
	number       = 1,
	pages        = {015005},
	doi          = {10.1088/2058-9565/aadd1a},
	note         = {Publisher: {IOP} Publishing},
	journaltitle = {Quantum Sci. Technol.},
	date         = {2018-09},
	langid       = {english}
}

@article{herr_lattice_2017,
	title        = {Lattice surgery translation for quantum computation},
	author       = {Herr, Daniel and Nori, Franco and Devitt, Simon J.},
	volume       = 19,
	number       = 1,
	pages        = {013034},
	doi          = {10.1088/1367-2630/aa5709},
	note         = {Publisher: {IOP} Publishing},
	journaltitle = {New J. Phys.},
	date         = {2017-01},
	langid       = {english}
}

@misc{gidney_magic_2024,
	title        = {Magic state cultivation: growing T states as cheap as {CNOT} gates},
	shorttitle   = {Magic state cultivation},
	author       = {Gidney, Craig and Shutty, Noah and Jones, Cody},
	publisher    = {{arXiv}},
	number       = {{arXiv}:2409.17595},
	doi          = {10.48550/arXiv.2409.17595},
	date         = {2024-09-26},
	eprinttype   = {arxiv},
	eprint       = {2409.17595}
}

@article{mcewen_relaxing_2023,
	title        = {Relaxing Hardware Requirements for Surface Code Circuits using Time-dynamics},
	author       = {{McEwen}, Matt and Bacon, Dave and Gidney, Craig},
	volume       = 7,
	pages        = 1172,
	doi          = {10.22331/q-2023-11-07-1172},
	note         = {Publisher: Verein zur Förderung des Open Access Publizierens in den Quantenwissenschaften},
	journaltitle = {Quantum},
	date         = {2023-11-07},
	langid       = {british}
}

@article{beverland_surface_2022,
	title        = {Surface Code Compilation via Edge-Disjoint Paths},
	author       = {Beverland, Michael and Kliuchnikov, Vadym and Schoute, Eddie},
	volume       = 3,
	number       = 2,
	pages        = {020342},
	doi          = {10.1103/PRXQuantum.3.020342},
	note         = {Publisher: American Physical Society},
	journaltitle = {{PRX} Quantum},
	date         = {2022-05-25}
}

@misc{debroy_luci_2024,
	title        = {{LUCI} in the Surface Code with Dropouts},
	author       = {Debroy, Dripto M. and {McEwen}, Matt and Gidney, Craig and Shutty, Noah and Zalcman, Adam},
	publisher    = {{arXiv}},
	number       = {{arXiv}:2410.14891},
	doi          = {10.48550/arXiv.2410.14891},
	date         = {2024-10-18},
	eprinttype   = {arxiv},
	eprint       = {2410.14891}
}

@INPROCEEDINGS{zhu_ecmas_2023,
  author={Zhu, Mingzheng and Fu, Hao and Wu, Jun and Zhang, Chi and Xie, Wei and Li, Xiang-Yang},
  booktitle={2024 IEEE/ACM International Symposium on Code Generation and Optimization (CGO)}, 
  title={Ecmas: Efficient Circuit Mapping and Scheduling for Surface Code}, 
  year={2024},
  volume={},
  number={},
  pages={158-169},
  doi={10.1109/CGO57630.2024.10444874}}

@InProceedings{silva_multi-qubit_2024,
  author =	{Silva, Allyson and Zhang, Xiangyi and Webb, Zak and Kramer, Mia and Yang, Chan-Woo and Liu, Xiao and Lemieux, Jessica and Chen, Ka-Wai and Scherer, Artur and Ronagh, Pooya},
  title =	{{Multi-qubit Lattice Surgery Scheduling}},
  booktitle =	{19th Conference on the Theory of Quantum Computation, Communication and Cryptography (TQC 2024)},
  pages =	{1:1--1:22},
  series =	{Leibniz International Proceedings in Informatics (LIPIcs)},
  ISBN =	{978-3-95977-328-7},
  ISSN =	{1868-8969},
  year =	{2024},
  volume =	{310},
  editor =	{Magniez, Fr\'{e}d\'{e}ric and Grilo, Alex Bredariol},
  publisher =	{Schloss Dagstuhl -- Leibniz-Zentrum f{\"u}r Informatik},
  address =	{Dagstuhl, Germany},
  URN =		{urn:nbn:de:0030-drops-206712},
  doi =		{10.4230/LIPIcs.TQC.2024.1}
}

@article{molavi_dependency-aware_2024,
  title = {Dependency-Aware Compilation for Surface Code Quantum Architectures},
  volume = {9},
  ISSN = {2475-1421},
  url = {http://dx.doi.org/10.1145/3720416},
  DOI = {10.1145/3720416},
  number = {OOPSLA1},
  journal = {Proceedings of the ACM on Programming Languages},
  publisher = {Association for Computing Machinery (ACM)},
  author = {Molavi,  Abtin and Xu,  Amanda and Tannu,  Swamit and Albarghouthi,  Aws},
  year = {2025},
  month = apr,
  pages = {57–84}
}

@article{cohen_low-overhead_2022,
	title        = {Low-overhead fault-tolerant quantum computing using long-range connectivity},
	author       = {Cohen, Lawrence Z. and Kim, Isaac H. and Bartlett, Stephen D. and Brown, Benjamin J.},
	volume       = 8,
	number       = 20,
	pages        = {eabn1717},
	doi          = {10.1126/sciadv.abn1717},
	journaltitle = {Sci. Adv.},
	date         = {2022-05-20},
	eprinttype   = {arxiv},
	eprint       = {2110.10794 [quant-ph]}
}

@article{vuillot_code_2019,
	title        = {Code deformation and lattice surgery are gauge fixing},
	author       = {Vuillot, Christophe and Lao, Lingling and Criger, Ben and Almudéver, Carmen García and Bertels, Koen and Terhal, Barbara M.},
	volume       = 21,
	number       = 3,
	pages        = {033028},
	doi          = {10.1088/1367-2630/ab0199},
	note         = {Publisher: {IOP} Publishing},
	journaltitle = {New J. Phys.},
	date         = {2019-03},
	langid       = {english}
}

@article{leroux_snakes_2024,
  title = {Snakes and Ladders: Adapting the Surface Code to Defects},
  volume = {6},
  ISSN = {2691-3399},
  url = {http://dx.doi.org/10.1103/815q-xjrb},
  DOI = {10.1103/815q-xjrb},
  number = {4},
  journal = {{PRX} Quantum},
  author = {Leroux,  Catherine and Lin,  Sophia F. and Bienias,  Przemyslaw and Sankar,  Krishanu R. and Benhemou,  Asmae and Kubica,  Aleksander and Iverson,  Joseph K.},
  year = {2025},
  pages = {040302} 
}

@article{bombin_topological_2006,
	title        = {Topological Quantum Distillation},
	author       = {Bombin, H. and Martin-Delgado, M. A.},
	volume       = 97,
	number       = 18,
	pages        = 180501,
	doi          = {10.1103/PhysRevLett.97.180501},
	journaltitle = {Phys. Rev. Lett.},
	date         = {2006-10-30},
	eprinttype   = {arxiv},
	eprint       = {quant-ph/0605138}
}

@misc{kobori_lsqca_2024-1,
	title        = {{LSQCA}: Resource-Efficient Load/Store Architecture for Limited-Scale Fault-Tolerant Quantum Computing},
	shorttitle   = {{LSQCA}},
	author       = {Kobori, Takumi and Suzuki, Yasunari and Ueno, Yosuke and Tanimoto, Teruo and Todo, Synge and Tokunaga, Yuuki},
	publisher    = {{arXiv}},
	number       = {{arXiv}:2412.20486},
	doi          = {10.48550/arXiv.2412.20486},
	date         = {2024-12-29},
	eprinttype   = {arxiv},
	eprint       = {2412.20486 [quant-ph]}
}

@inproceedings{shor1996fault,
  title={Fault-tolerant quantum computation},
  author={Shor, Peter W},
  booktitle={Proceedings of 37th conference on foundations of computer science},
  pages={56--65},
  year={1996},
  organization={IEEE},
    doi = {10.1109/SFCS.1996.548464}
}

@article{preskill1998reliable,
  title={Reliable quantum computers},
  author={Preskill, John},
  journal={Proceedings of the Royal Society of London. Series A: Mathematical, Physical and Engineering Sciences},
  volume={454},
  number={1969},
  pages={385--410},
  year={1998},
  publisher={The Royal Society},
    doi={10.1098/rspa.1998.0167}
}

@article{steane1999efficient,
  title={Efficient fault-tolerant quantum computing},
  author={Steane, Andrew M},
  journal={Nature},
  volume={399},
  number={6732},
  pages={124--126},
  year={1999},
  publisher={Nature Publishing Group UK London}
}

\end{document}